\documentclass[12pt, oneside, reqno]{amsart}
 \usepackage{graphicx}
\usepackage{amsmath,amsthm,amsfonts,amscd,amssymb,mathrsfs,hyperref}
\usepackage[numeric,lite,initials]{amsrefs}
\usepackage[all]{xypic}
\usepackage{amscd}   
\usepackage[all]{xy} 
\usepackage{booktabs} 
\usepackage{tikz}
\usetikzlibrary{babel}
\usepackage{physics}
\usepackage{nccmath} 
\DeclareMathAlphabet{\mathscr}{OT1}{pzc}{m}{it} 
\makeatletter
\g@addto@macro\normalsize{%
  \setlength\abovedisplayskip{4pt}
  \setlength\belowdisplayskip{4pt}
  \setlength\abovedisplayshortskip{4pt}
  \setlength\belowdisplayshortskip{4pt}
}
\makeatother
\usepackage{hyperref}
  \hypersetup{colorlinks=true,citecolor=blue, urlcolor= cyan}

 \AtBeginDocument{%
    \def\MR#1{}
 }

\usepackage{color}

\newtheorem{theorem}{Theorem}[section]

\newcommand{\isom}{\cong}

\newcommand{\SL}{\operatorname{SL}}

\newcommand{\PP}{\mathbb{P}}

\newcommand{\CC}{\mathbb{C}}
\newcommand{\ZZ}{\mathbb{Z}}

\newcommand{\Seg}{\operatorname{Seg}}

\def\phi{ \varphi }

\def \g{\mathfrak{g}}

\theoremstyle{definition}

\theoremstyle{remark}
\newtheorem{remark}[theorem]{Remark}

\newcommand{\Spin}{\operatorname{Spin}}
\newcommand{\HDet}{\operatorname{HDet}}

\begin{document}
\date{\today}

\author{Fr\'ed\'eric Holweck}\email{frederic.holweck@utbm.fr}
\address{Laboratoire Interdisciplinaire Carnot de Bourgogne, ICB/UTBM, UMR 6303 CNRS,
Universit\'e Bourgogne Franche-Comt\'e, 90010 Belfort Cedex, France
}
\author{Luke Oeding}\email{oeding@auburn.edu}
\address{Department of Mathematics and Statistics,
Auburn University,
Auburn, AL, USA
}

\title{A hyperdeterminant on Fermionic Fock Space}

\begin{abstract}
Twenty years ago Cayley's hyperdeterminant, the degree four invariant of the polynomial ring $\CC[\CC^2\otimes\CC^2\otimes \CC^2]^{{\SL_2(\CC)}^{\times 3}}$, was popularized in modern physics as separates genuine entanglement classes in the three qubit Hilbert space and is connected to entropy formulas for special solutions of black holes. In this note we compute the analogous invariant on the fermionic Fock space for $N=8$, i.e. spin particles with four different locations, and show how this invariant projects to other well-known invariants in quantum information. We also give combinatorial interpretations of these formulas. 
\end{abstract}

\maketitle

\section{Introduction}
Arthur Cayley in 1845 \cite{cayley1800} established several notions of hyperdeterminant as possible analogues of the classical notion of determinant but for hypermatrices. The most popular hyperdeterminant (see \cite{GKZ}), denoted $\HDet$, arises by generalizing the concept of singular matrix to hypermatrix. Let $A=(a_{ijk})_{i,j,k\in\{0,1\}}$ be a real/complex $2\times  2\times 2$ tensor. The hyperdeterminant is:

\begin{multline*}
\HDet_{222}(A) = a_{000}^{2}a_{111}^{2}+a_{010}^{2}a_{101}^{2}+a_{001}^{2}a_{110}^{2}+a_{011}^{2}a_{100}^{2} +4(a_{000}a_{011}a_{101}a_{110}+a_{001}a_{010}a_{100}a_{111})\\
-2(a_{000}a_{001}a_{110}a_{111}+a_{000}a_{010}a_{101}a_{111}+a_{000}a_{011}a_{100}a_{111}+\\
a_{001}a_{010}a_{101}a_{110}+a_{001}a_{011}a_{100}a_{110}+a_{010}a_{011}a_{100}a_{101})
.\end{multline*}

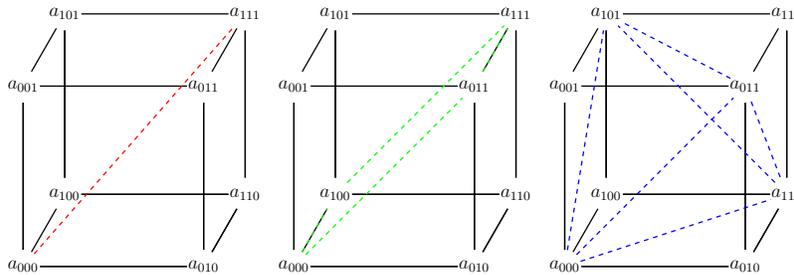
\begin{figure}[!h]
 \centering
 \scalebox{0.6}
 {
 \begin{tikzpicture}[scale=4]
	 \tikzstyle{vertex}=[circle,minimum size=20pt,inner sep=0pt]
	 \tikzstyle{selected vertex} = [vertex, fill=red!]
	 \tikzstyle{selected edge} = [draw,dashed,thick,red!]
	 \tikzstyle{selected edge2} = [draw,dashed,thick,green!]
	 \tikzstyle{selected edge3} = [draw,dashed,thick,blue!]
	 \tikzstyle{edge} = [draw,thick,-,black]
	 \node[vertex] (v0) at (0,0) {$a_{000}$};
	 \node[vertex] (v1) at (0,1) {$a_{001}$};
	 \node[vertex] (v2) at (1,0) {$a_{010}$};
	 \node[vertex] (v3) at (1,1) {$a_{011}$};
	 \node[vertex] (v4) at (0.23, 0.4) {$a_{100}$};
	 \node[vertex] (v5) at (0.23,1.4) {$a_{101}$};
	 \node[vertex] (v6) at (1.23,0.4) {$a_{110}$};
	 \node[vertex] (v7) at (1.23,1.4) {$a_{111}$};
	 \draw[edge] (v0) -- (v1) -- (v3) -- (v2) -- (v0);
	 \draw[edge] (v0) -- (v4) -- (v5) -- (v1) -- (v0);
	 \draw[edge] (v2) -- (v6) -- (v7) -- (v3) -- (v2);
	 \draw[edge] (v4) -- (v6) -- (v7) -- (v5) -- (v4);
	 \draw[edge] (v0) -- (v2);
	 \draw[edge] (v2) -- (v6);
	 \draw[edge] (v6) -- (v4);
	 \draw[edge] (v4) -- (v5);	
	 \draw[selected edge] (v0) -- (v7);
	 \node[vertex] (vv0) at (1.5,0) {$a_{000}$};
	 \node[vertex] (vv1) at (1.5,1) {$a_{001}$};
	 \node[vertex] (vv2) at (2.5,0) {$a_{010}$};
	 \node[vertex] (vv3) at (2.5,1) {$a_{011}$};
	 \node[vertex] (vv4) at (1.73, 0.4) {$a_{100}$};
	 \node[vertex] (vv5) at (1.73,1.4) {$a_{101}$};
	 \node[vertex] (vv6) at (2.73,0.4) {$a_{110}$};
	 \node[vertex] (vv7) at (2.73,1.4) {$a_{111}$};
	 \draw[edge] (vv0) -- (vv1) -- (vv3) -- (vv2) -- (vv0);
	 \draw[edge] (vv0) -- (vv4) -- (vv5) -- (vv1) -- (vv0);
	 \draw[edge] (vv2) -- (vv6) -- (vv7) -- (vv3) -- (vv2);
	 \draw[edge] (vv4) -- (vv6) -- (vv7) -- (vv5) -- (vv4);
	 \draw[edge] (vv0) -- (vv2);
	 \draw[edge] (vv2) -- (vv6);
	 \draw[edge] (vv6) -- (vv4);
	 \draw[edge] (vv4) -- (vv5);	
	 \draw[selected edge2] (vv0) -- (vv4);
	 \draw[selected edge2] (vv3) -- (vv7);
	 \draw[selected edge2] (vv0) -- (vv3);
	 \draw[selected edge2] (vv4) -- (vv7);
	 \node[vertex] (vvv0) at (3,0) {$a_{000}$};
	 \node[vertex] (vvv1) at (3,1) {$a_{001}$};
	 \node[vertex] (vvv2) at (4,0) {$a_{010}$};
	 \node[vertex] (vvv3) at (4,1) {$a_{011}$};
	 \node[vertex] (vvv4) at (3.23, 0.4) {$a_{100}$};
	 \node[vertex] (vvv5) at (3.23,1.4) {$a_{101}$};
	 \node[vertex] (vvv6) at (4.23,0.4) {$a_{110}$};
	 \node[vertex] (vvv7) at (4.23,1.4) {$a_{111}$};
	 \draw[edge] (vvv0) -- (vvv1) -- (vvv3) -- (vvv2) -- (vvv0);
	 \draw[edge] (vvv0) -- (vvv4) -- (vvv5) -- (vvv1) -- (vvv0);
	 \draw[edge] (vvv2) -- (vvv6) -- (vvv7) -- (vvv3) -- (vvv2);
	 \draw[edge] (vvv4) -- (vvv6) -- (vvv7) -- (vvv5) -- (vvv4);
	 \draw[edge] (vvv0) -- (vvv2);
	 \draw[edge] (vvv2) -- (vvv6);
	 \draw[edge] (vvv6) -- (vvv4);
	 \draw[edge] (vvv4) -- (vvv5);	
	 \draw[selected edge3] (vvv0) -- (vvv6);
	 \draw[selected edge3] (vvv0) -- (vvv3);
	 \draw[selected edge3] (vvv0) -- (vvv5);
	 \draw[selected edge3] (vvv3) -- (vvv6);
	 \draw[selected edge3] (vvv3) -- (vvv5);
	 \draw[selected edge3] (vvv6) -- (vvv5);
 \end{tikzpicture}
 }
 \caption{Monomials of Cayley's Hyperdeterminant from a combinatorial perspective: The four diagonals (red) of the cube provide the first four monomials of degree $4$ of type $a_{ijk}a_{\overline{ijk}}$ (where bar denotes the bit-complement), the six parallelograms (green) provide the six monomials of type $a_{ijk}a_{\overline{ijk}}a_{i'j'k'}a_{\overline{i'j'k'}}$ (where $ijk$ and $i'j'k'$ have one bit difference), and the two tetrahedra (blue) give the two monomials of type $a_{ijk}a_{i\overline{jk}}a_{\overline{i}j\overline{k}}a_{\overline{ij}k}$.}\label{fig:hdet}
 \end{figure}

There is a combinatorial picture associated to this polynomial \cite{bremner2012}: Consider the  cube of Figure~\ref{fig:hdet} labelled by the entries of the $2\times 2\times 2$ tensor $A$, the three groups of $\HDet_{222}(A)$ monomials are deduced from the diagonals, parallelograms and tetrahedra inside the cube.
 
 This polynomial gained a lot of attention in the Quantum Information literature in the early $2000's$ when Miyake \cite{miyake_hyperdet} showed that this invariant is useful to distinguish the different types of genuine three-qubit entanglement. In the STU model, extremal back holes solutions with 4 electric and 4 magnetic charges have entropy formula that is the square root of Cayley's hyperdeterminant \cite{duff2007}. This surprising analogy was the beginning of several works on the black holes/qubits correspondence \cite{borsten2009,borsten2012,kallosh2006,levay_2006,levay_2008}.
 From an algebraic geometry perspective $\HDet=0$ is the equation of the dual variety $X^\vee$ of the Segre embedding of three copies of $\PP^1$ in $\PP^7$, i.e. $X=\Seg(\PP^1\times\PP^1\times \PP^1)\subset \PP(\CC^2\otimes\CC^2\otimes \CC^2)$. Recall that if $X\subset \PP(V)$ is a projective algebraic variety, i.e. the zero set of a collection of homogeneous polynomials, then 
 \begin{equation}
  X^\vee=\overline{\{H\in \PP(V^*), \exists x\in X_{\text{smooth}}, T_x X\subset H\}}.
 \end{equation}\label{eq:dual}
In other words, the dual variety of $X$, is the variety (in the dual projective space) of tangent hyperplanes, i.e. hyperplanes that intersect $X$ tangentially. The bar in~\ref{eq:dual} refers to the Zariski closure, i.e. the minimal set defined by polynomial equations that contains the set. When $X^\vee$ is a hypersurface, its defining equation, $\Delta_X$, is called the $X$-discriminant \cite{GKZ}. When $X=\Seg(\PP^{d_1}\times\PP^{d_2}\times \dots\times \PP^{d_n})\subset \PP(\CC^{d_1+1}\otimes \CC^{d_2+1}\otimes \dots\otimes \CC^{d_n+1})$ and $d_i\leq \sum_{j\neq i} d_j$ then the $X$-discriminant is always a hypersurface called Hyperdeterminant of format $(d_1+1)\times (d_2+1)\times\dots\times (d_n+1)$ and denoted by $\HDet_{d_1+1,d_2+1,\dots,d_n+1}$. Cayley's hyperdeterminant is the hyperdeterminant of format $2\times 2 \times 2$.

In this paper we consider the analogue of Cayley's hyperdeterminant in the $\mathcal{F}^+$ component of the fermionic Fock spaces $\mathcal{F}_{2N}$, i.e. $\mathcal{F}_{2N}=\bigwedge ^{\bullet} V=\bigoplus_{k \text{even}} \bigwedge ^k V \oplus\bigoplus_{k \text{odd}} \bigwedge ^k V =\mathcal{F}^+\oplus \mathcal{F}^-$ where $V$ is a $N$-dimensional single particle space. We use techniques established by the authors in \cite{HolweckOeding} to compute it on a generic element of $\mathcal{F}^+$ in the case where the single particle state of the Fock space is $8$-dimensional, spin $\frac{1}{2}$-particle in $4$ different locations, i.e. $V=\CC^8=\underbrace{\CC^4}_{\text{locations}}\otimes \underbrace{\CC^2}_{\Spin}$. We show that this polynomial contains copies of the Cayley analogue for $4$ fermions with $8$-single particle states, the $4$-qubit hyperdeterminant and the $4$-bosonic qubit discriminant.

The paper is organized as follows. In Section~\ref{sec:fock} we recall the principle of the $\Spin(2N)$ action on the fermionic Fock space $\mathcal{F}_{2N}$ and why this action is the natural generalization of the SLOCC group on multiqubit Hilbert space. This allows to connect entanglement classification in fermionic Fock space and classification of spinors as explained in \cite{SarosiLevay,LevayHolweck}. In Section~\ref{sec:e8} we compute the equation of the dual variety of $X_{\Spin(16)}\subset \PP(\mathcal{F}^+)$ on a generic element of $\mathcal{F}^+$. Our computation is based on techniques of \cite{HolweckOeding} and the realization of the exceptional Lie algebra as a $\ZZ_2$-graded algebra $\mathfrak{e}_8=\mathfrak{s}\mathfrak{o}(16)\oplus \mathcal{F}^+$. This polynomial of degree $240$ in $8$ variables has a nice combinatorial presentation as planes of a finite geometric cube space. It contains $8$ copies of the degree $120$ polynomial in $7$ variables that measures entanglement in $\mathcal{H}=\bigwedge^4 \CC^8$ and several copies of the four-qubit hyperdeterminant. These combinatorial pictures are provided in Section~\ref{sec:fano}. Section~\ref{sec:conclusion} is dedicated to concluding remarks.
\section{Entanglement in fermionic Fock space}\label{sec:fock}
Here we recall how the fermionic Fock space of a $2n$-single particle state Hilbert space  can seen as a Hilbert space with an SLOCC action given by the spin group $\Spin(2n,\CC)$ \cite{SarosiLevay,LevayHolweck}.

Let $\mathcal{H}=\CC^N$ denote a Hilbert space for $N$-single particle states with $N=2n$ and denote by $\mathcal{F}=\bigwedge ^*\mathcal{H}$ the fermionic Fock space obtained from $\mathcal{H}$. We equip $\mathcal{H}$ with a canonical basis $\{e_I\}$ and we denote by $\{e^J\}$  the basis of the dual space $\mathcal{H}^*$. Consider the vector space $\mathcal{V}=\mathcal{H}\oplus\mathcal{H}^*$ where vectors are elements $x=v+\alpha$ with $v\in \mathcal{H}$ and $\alpha\in\mathcal{H}^*$. Let us equip $\mathcal{V}$ with the quadratic form 
$g=\left(\begin{smallmatrix}
              0 & I_N\\
              I_N & 0\end{smallmatrix}\right)$,
which makes $\mathcal{H}$ and $\mathcal{H}^*$ orthogonal complementary subspaces in $\mathcal{V}$. For all $x\in \mathcal{V}$ one defines an operator acting on  $\mathcal{F}=\bigwedge^*\mathcal{H}$ as follows: Consider the operators $\hat{e}_I$ and $\hat{e}^J$ corresponding to the exterior and interior products, i.e.
              \begin{equation}
\begin{matrix}
               \hat{e}_I \colon & \mathcal{F} & \to & \mathcal{F} \\
                         & f & \mapsto & \sqrt{2}e_I\wedge f                
\end{matrix}\quad, \quad \quad \quad
\begin{matrix} \hat{e}^J \colon & \mathcal{F} & \to & \mathcal{F}\\
               &  f & \mapsto & \sqrt{2}e_J\lrcorner f
\end{matrix}\quad
.
              \end{equation}
Then to $x=v^Ie_I+\alpha_Je^J\in \mathcal{V}$ (summation over repeated indices) one can associate the operator $\mathcal{O}_x=v^I\hat{e}_I+\alpha_J\hat{e}^J$ that acts on $\mathcal{F}$. The operators $\hat{e}_I$ and $\hat{e}^J$ are respectively known as creation and annihilation operators and will be denoted from now on by $p_I$ and $n_J$. Let $\ket{0}$ denote a unit vector of $\CC=\wedge^0 \mathcal{H}$ defined by the property that $n_J\ket{0}=0$, $J=1,\dots,N$. The state $\ket{0}$ is known as the vacuum and corresponds to a state without any excitations. Then an arbitrary state of $\mathcal{F}$ can be expressed as 
\begin{equation}\label{eq:femionicstate}
 \ket{\Psi}=\sum_{m=0} ^N \sum_{I_1,\dots,I_m=1}^N \psi^m_{I_1,\dots,I_m}p_{I_1}\dots p_{I_m}\ket{0},
\end{equation}
where $\psi^m_{I_1,\dots,I_m}$ are  skew-symmetric tensors. In the second quantization picture it translates the idea that quantum states may be obtained from the vacuum by excitations.

Note that the creation and annihilation operators, $p_I$ and $n_J$ satisfy the Canonical Anticommutation Relations (CAR):
\begin{equation}
 \begin{array}{ccc}
  \{p_I,p_J\}=0 & \{n_I,n_J\}=0 & \{p_I,n_J\}=2\delta_I^J,
 \end{array}
\end{equation}
where $\{\cdot,\cdot \}$ is the standard anti-commutator. 
From the CAR one can define a product on $\mathcal{V}$ such that $xy=z$ for $x,y, z\in \mathcal{V}$, where $z$ is such that $\mathcal{O}_x\mathcal{O}_y=\mathcal{O}_z$. On can easily check that given $x\in \mathcal{V}$ one has 
\begin{equation}
x^2=Q(x,x){\bf 1},
\end{equation} 
where $Q$ is the quadratic form on $\mathcal{V}$ defined by $g$. 
This shows that $\mathcal{V}$ with this product is the Clifford algebra $\mathcal{C}(\mathcal{V},Q)$. This algebra acts on $\mathcal{F}$ by sending $x\mapsto \mathcal{O}_x$.

Recall that the existence of $g$ allows us to define the Lie algebra $\mathfrak{s}\mathfrak{o}(\mathcal{V},Q)=\mathfrak{s}\mathfrak{o}(2N,\CC)$ as the set of matrices $s\in \mathcal{M}_{2N\times 2N}(\CC)$ such that,
\begin{equation}\label{eq:smatrix}
 s=\begin{pmatrix}
    A & B\\
    C & -^tA
   \end{pmatrix},
\end{equation}
where $B$ and $C$ satisfy $B=-^tB$ and $C=-^tC$. In other words, $s$ is a skew symmetric matrix, i.e. $\mathfrak{s}\mathfrak{o}(\mathcal{V},Q)\isom \bigwedge^2\mathcal{V}$. The Lie algebra $\mathfrak{s}\mathfrak{o}(\mathcal{V},Q)$ can be embedded into $\mathcal{C}(\mathcal{V},Q)$ via,
\begin{equation}\label{eq:embedding}
 \begin{array}{ccc}
  \bigwedge^2\mathcal{V} & \to & \mathcal{C}(\mathcal{V},Q)\\
  x\wedge y & \mapsto & \frac{1}{4}[x,y].
 \end{array}
\end{equation}
This embedding defines a representation of $\mathfrak{s}\mathfrak{o}(\mathcal{V},Q)$ acting on $\mathcal{F}$. Indeed Eq.~\eqref{eq:embedding} maps $s$  (from Eq.~\eqref{eq:smatrix}) to the operator $\mathcal{O}_s$ where
\begin{equation}\label{eq:spinaction}
 \mathcal{O}_s=\frac{1}{2} \sum_{i,j} A_{ij}[p_i,n_j]+B_{ij}p_ip_j+C_{ij}n_in_j.
\end{equation}

The action of Eq.~\eqref{eq:spinaction} preserves the parity of the number of creations  and annihilators needed to describe $\ket{\Psi}$ from the vacuum Eq.~\eqref{eq:femionicstate}, i.e. one has a decomposition of $\mathcal{F}_{N}$ in two irreducible representations,
\begin{equation}
 \mathcal{F}_{N}=\mathcal{F}_{N}^+\oplus \mathcal{F}_{N}^{-}.
\end{equation}

Here $\mathcal{F}_{N} ^+$ (resp. $\mathcal{F}_{N} ^{-}$) denotes the fermionic states obtained by applying an even (resp. odd) number of creation operators to the vacuum.

The action of $\mathfrak{s}{\mathfrak{o}}(\mathcal V,2N)$ on $\mathcal{F}_{N}^{\pm}$ is known in the mathematics literature as the spin representation \cite{FultonHarris}. The Lie group with Lie algebra $\mathfrak{s}{\mathfrak{o}}(\mathcal V,2N)$ acting on $\mathcal{F}_{N}^{\pm}$ is the spin group, $\Spin(2N,\CC)$, i.e. the double covering of $\operatorname{SO}(2N,\CC)$. The Spin group has a unique closed orbit on $\PP(\mathcal{F}_N ^{\pm})$ known in representation theory as the highest weight orbit or the variety of pure spinors (i.e. obtained from $\ket{0}$ by the action of the spin group). In this setting it is natural to consider as separable fermionic Fock states \cite{SarosiLevay} the ones obtained from the vacuum by reversible operations of the SLOCC group $\Spin(2n,\CC)$, i.e.
\begin{equation}
 X_{\text{sep}}=\PP(\Spin(2n,\CC)\cdot\ket{0})\subset \PP(\mathcal{F}_N ^{\pm}).
\end{equation}

In \cite{LevayHolweck} it was shown that for $N=2n$ fermionic, bosonic and multiqubit systems ($n$-fermions with $2n$-single particles states, $n$-bosonic qubits, $n$-qubits) can be embedded in many ways in $\mathcal{F}_{N}^{\pm}$. Under these embeddings, the $\Spin(2N,\CC)$ action on $\mathcal{F}_{N}^{\pm}$ boils down to the usual SLOCC groups on the corresponding embedded Hilbert spaces ($\SL(n,\CC)$, $\SL(2,\CC)$, $\SL(2,\CC)^{\times n}$). In this respect it is natural to look for Spin invariant polynomials in order to separate entanglement classes in fermionic Fock space.

\section{The hyperdeterminant on fermionic Fock space for $N=8$}\label{sec:e8}
We now focus on the expression of the equation of the dual of $X_{\text{sep}}$ i.e. the analogue of Cayley's hyperdeterminant for $\Spin(16,\CC)$ and $\mathcal{F}_8^{+}$.
In \cite{HolweckOeding} we showed how dual equations of orbit closures a Lie group $G$ acting on its Lie algebra $\mathfrak{g}$ can be projected to submodules by restriction. Let us recall the principle of our construction. Let $\mathfrak{g}_0$ denote a subalgebra of $\mathfrak{g}$ acting on a module $\mathfrak{g}_1$ such that as vector spaces we have
\begin{equation}\label{eq:z2}
 \mathfrak{g}=\mathfrak{g}_0\oplus\mathfrak{g}_1.
\end{equation}
The Lie bracket on $\mathfrak{g}$ is compatible with the bracket on $\mathfrak{g}_0$ and respects the $\ZZ_2$ grading at \eqref{eq:z2}. More specifically, the action of $\mathfrak{g}_0$ on $\mathfrak{g}_1$ defines the bracket $[\cdot ,\cdot ] \colon \g_0 \times \g_1 \to \g_1$, and we can insist that it is skew commuting, and there is a consistent restriction of the bracket of $\g$ to $\mathfrak{g}_1$ i.e. $[\cdot ,\cdot ]:\mathfrak{g}_1\times\mathfrak{g}_1\to \mathfrak{g}_0$ (see \cite{HolweckOeding2}).
Let $X_G\subset \PP(\mathfrak{g})$ denote the unique closed $G$-orbit also known as the adjoint variety for the Lie group $G$ and $Y_{G_0}\subset \PP(\mathfrak{g}_1)$ the unique closed $G_0$-orbit in $\PP(\mathfrak{g}_1)$ known as the highest weight orbit of the $\mathfrak{g}_1$-module for the Lie group $G_0$. With these notations, in \cite[Theorem 3.1]{HolweckOeding} we showed that $Y_{G_0}^\vee\subset X_G ^\vee\cap \PP(\mathfrak{g}_1)$. In particular if  $Y_{G_0}^\vee$ is a hypersurface, its defining equation $\Delta_{Y_{G_0}}$ will divide the restriction to $\PP(\mathfrak{g}_1)$ of $\Delta_{G}$, the defining equation of $X_G^\vee$. When the polynomials have the same degree, one gets equality.

\begin{remark}
 Theorem 3.1 of \cite{HolweckOeding} is in fact more general and deals with $\ZZ_k$-graded Lie algebras. Recently Manivel and Benedetti \cite{Manivel} improved our result by relating the degree relations between the polynomials $\Delta_{Y_{G_0}}$ and $\Delta_{X_G}$ with the grading of $\mathfrak{g}$.
\end{remark}
Let us consider the following realization of the exceptional Lie algebra $\mathfrak{e}_8$,
\begin{equation}
 \mathfrak{e}_8=\mathfrak{s}\mathfrak{o}(16,\CC)\oplus \mathcal{F}_8^{+}.
\end{equation}
This $\ZZ_2$-branching of $\mathfrak{e}_8$ with $\mathfrak{g}_1=\mathcal{F}_8^+$ was used by Antonyan and Elashvili \cite{antonyan1982} to obtain the orbit classification of spinors in $\mathcal{F}_{8} ^{\pm}$. The bracket on $ \mathfrak{s}\mathfrak{o}(16,\CC)$ is the usual bracket on this Lie algebra, while the bracket $[\cdot, \cdot]:\mathfrak{g}_0\times \mathfrak{g}_1\to \mathfrak{g}_1$ is defined by Eq.~\eqref{eq:spinaction}. The bracket $[\cdot, \cdot]:\mathfrak{g}_1\times\mathfrak{g}_1\to \mathfrak{g}_0$ is described in \cite[Eqs.~(8) and (9)]{antonyan1982}, and essentially reflects the Jordan algebra structure.

Our strategy to find an expression of $X_{\text{sep}}^\vee$ will be to restrict an expression of $X_{E_8}^\vee$, which we need to construct. Such an expression exists on a Cartan subalgebra of $\mathfrak{e}_8$, i.e. a subspace  of semi-simple elements of dimension $8$. Recall that the Chevalley's restriction Theorem insures that for a Lie algebra $\mathfrak{g}$ with Cartan algebra $\mathfrak{h}$ then $\CC[\mathfrak{g}]^G\simeq \CC[\mathfrak{h}]^W$ where $W$ is the Weyl group of type $G$. Under this restriction, Tevelev \cite{Tevelev}  showed that when $G$ is a simple Lie group with simply laced Dynkin diagram that 
\[ \Delta_G \mapsto \Pi_{\alpha\in R} \alpha\in \CC[\mathfrak{h}]^W,\]
 where $R$ is the set of roots of $\mathfrak{g}$. In other words, the restriction of the $G$-discriminant of a Lie algebra to semi-simple elements of $\mathfrak{g}$ is the product of the roots. In order to restrict $\Delta_G$ to $\mathfrak{g}_1=\mathcal{F}_8^+$ one needs to identify a Cartan subalgebra of $\mathfrak{e}_8$ living inside $\mathcal{F}_8^{+}$. Recall that a standard choice \cite{antonyan1982} for  a Cartan of $\mathfrak{e}_8$ is $\mathfrak{h}=\{\sum_{i=1}^8 x_i e_{i}\otimes e^i-e_{i+8}\otimes e^{i+8}, x_i\in \CC\}$. For this Cartan the $240$ roots of $\mathfrak{e}_8$ are
\begin{equation}\label{eq:Roots}
\pm x_i\pm x_j, \frac{1}{2}(\pm x_1\pm x_2\pm x_3\pm x_4\pm x_5\pm x_6\pm x_7\pm x_8)
\end{equation}
with an even number of minus signs for the second type of roots. A generic element of a Cartan subalgebra $\mathfrak{c}$, such that  $\mathfrak{c}\subset \mathcal{F}_8^+$ can be described \cite{LevayHolweck} as,

\begin{equation}\label{eq:semisimple}
 \ket{\Psi}=\sum_{i=1}^8 y_i \ket{E_i},
\end{equation}
where we have chosen the following basis of $\mathfrak{c}$:
\begin{equation}
 \begin{array}{cc}
  \ket{E_1}=(p_1p_2p_3p_4+p_{\overline{1}}p_{\overline{2}}p_{\overline{3}}p_{\overline{4}})\ket{0} & \ket{E_2}=(p_1p_2p_{\overline{3}}p_{\overline{4}}+p_{\overline{1}}p_{\overline{2}}p_3p_4)\ket{0}\\[.5ex]
  \ket{E_3}=(p_1p_{\overline{2}}p_3p_{\overline{4}}+p_{\overline{1}}p_{{2}}p_{\overline{3}}p_{{4}})\ket{0} & \ket{E_4}=(p_1p_{\overline{2}}p_{\overline{3}}p_{{4}}+p_{\overline{1}}p_{{2}}p_3p_{\overline{4}})\ket{0}\\[.5ex]
   \ket{E_5}=(p_1p_{\overline{1}}p_4p_{\overline{4}}+p_2p_{\overline{2}}p_3p_{\overline{3}})\ket{0} &  \ket{E_6}=(p_1p_{\overline{1}}p_3p_{\overline{3}}+p_2p_{\overline{2}}p_4p_{\overline{4}})\ket{0}\\[.5ex]
   \ket{E_7}=(p_1p_{\overline{1}}p_2p_{\overline{2}}+p_2p_{\overline{2}}p_4p_{\overline{4}})\ket{0} & \ket{E_8}=({\bf 1}+p_1p_2p_3p_4p_{\overline{1}}p_{\overline{2}}p_{\overline{3}}p_{\overline{4}})\ket{0}.
 \end{array}
\end{equation}

In order to obtain the projection of $\Delta_G$ to the semi-simple algebra $\mathfrak{c}$ one needs to express the variables $x_i$ in terms of $y_j$. This can be achieved by considering \cite{LevayHolweck}:
\begin{equation}\label{eq:yx}
\scalebox{.85}{$ \begin{array}{cc}
  y_1=\frac{1}{2}(x_1+x_2+x_3+x_4-x_5-x_6-x_7-x_8), & y_2=\frac{1}{2}(x_1+x_2-x_3-x_4-x_5-x_6+x_7+x_8)\\[.5ex]
  y_3=\frac{1}{2}(x_1-x_2+x_3-x_4-x_5+x_6-x_7+x_8), & y_4=\frac{1}{2}(x_1-x_2-x_3+x_4-x_5+x_6+x_7-x_8)\\[.5ex]
   y_5=\frac{1}{2}(x_1-x_2-x_3+x_4+x_5-x_6-x_7+x_8), & y_6=\frac{1}{2}(x_1-x_2+x_3-x_4+x_5-x_6+x_7-x_8)\\[.5ex]
  y_7=\frac{1}{2}(x_1+x_2-x_3-x_4+x_5+x_6-x_7-x_8), & y_8=\frac{1}{2}(x_1+x_2+x_3+x_4+x_5+x_6+x_7+x_8)
 \end{array}
$ }
\end{equation}
The fundamental invariants can be expressed in terms of the $y_{i}$'s by inverting the relations above, writing the set of roots $R$ in the $y$-basis and forming the power sums
\[f_{d} = \sum_{\alpha\in R} \alpha^{d},
\] for $d$ in $\{2,8,12,14,18,20, 24,30\}$. We did this computation in Macaulay2 \cite{M2}. 
Up to symmetry permuting the names of the variables and rescaling the invariant, the monomials that occur in each are listed in the appendix.

The expression of the roots at \eqref{eq:Roots} after the substitution from the inverse of \eqref{eq:yx} has every root paired with its negative. As such the restriction $\Delta_{E_8|\mathfrak{c}}$, which is the expression of the $\HDet_{\Spin(16,\CC)}$ of $\mathcal{F}_8^{+}$ on semi-simple elements (Eq.~\eqref{eq:semisimple}) becomes (after re-scaling)
\begin{equation}\label{eq:hdetspin}
\begin{array}{l}
\text{HDet}_{\Spin(16,\CC)}(\Psi)=
\Bigl(y_{8}y_{7}y_{6}y_{5}y_{4}y_{3}y_{2}y_{1}\\
(y_{5}-y_{6}-y_{7}-y_{8})(y_{5}-y_{6}-y_{7}+y_{8})(y_{5}-y_{6}+y_{7}-y_{8})(y_{5}-y_{6}+y_{7}+y_{8})\\
(y_{5}+y_{6}-y_{7}-y_{8})(y_{5}+y_{6}-y_{7}+y_{8})(y_{5}+y_{6}+y_{7}-y_{8})(y_{5}+y_{6}+y_{7}+y_{8})\\
(y_{3}-y_{4}-y_{7}-y_{8})(y_{3}-y_{4}-y_{7}+y_{8})(y_{3}-y_{4}+y_{7}-y_{8})(y_{3}-y_{4}+y_{7}+y_{8})\\
(y_{3}-y_{4}-y_{5}-y_{6})(y_{3}-y_{4}-y_{5}+y_{6})(y_{3}-y_{4}+y_{5}-y_{6})(y_{3}-y_{4}+y_{5}+y_{6})\\
(y_{3}+y_{4}-y_{7}-y_{8})(y_{3}+y_{4}-y_{7}+y_{8})(y_{3}+y_{4}+y_{7}-y_{8})(y_{3}+y_{4}+y_{7}+y_{8})\\
(y_{3}+y_{4}-y_{5}-y_{6})(y_{3}+y_{4}-y_{5}+y_{6})(y_{3}+y_{4}+y_{5}-y_{6})(y_{3}+y_{4}+y_{5}+y_{6})\\
(y_{2}-y_{4}-y_{6}-y_{8})(y_{2}-y_{4}-y_{6}+y_{8})(y_{2}-y_{4}+y_{6}-y_{8})(y_{2}-y_{4}+y_{6}+y_{8})\\
(y_{2}-y_{4}-y_{5}-y_{7})(y_{2}-y_{4}-y_{5}+y_{7})(y_{2}-y_{4}+y_{5}-y_{7})(y_{2}-y_{4}+y_{5}+y_{7})\\
(y_{2}+y_{4}-y_{6}-y_{8})(y_{2}+y_{4}-y_{6}+y_{8})(y_{2}+y_{4}+y_{6}-y_{8})(y_{2}+y_{4}+y_{6}+y_{8})\\
(y_{2}+y_{4}-y_{5}-y_{7})(y_{2}+y_{4}-y_{5}+y_{7})(y_{2}+y_{4}+y_{5}-y_{7})(y_{2}+y_{4}+y_{5}+y_{7})\\
(y_{2}-y_{3}-y_{6}-y_{7})(y_{2}-y_{3}-y_{6}+y_{7})(y_{2}-y _{3}+y_{6}-y_{7})(y_{2}-y_{3}+y_{6}+y_{7})\\
(y_{2}-y_{3}-y_{5}-y_{8})(y_{2}-y_{3}-y_{5}+y_{8})(y_{2}-y_{3}+y_{5}-y_{8})(y_{2}-y_{3}+y_{5}+y_{8})\\
(y_{2}+y_{3}-y_{6}-y_{7})(y_{2}+y_{3}-y_{6}+y_{7})(y_{2}+y_{3}+y_{6}-y_{7})(y_{2}+y_{3}+y_{6}+y_{7})\\
(y_{2}+y_{3}-y_{5}-y_{8})(y_{2}+y_{3}-y_{5}+y_{8})(y_{2}+y_{3}+y_{5}-y_{8})(y_{2}+y_{3}+y_{5}+y_{8})\\
(y_{1}-y_{4}-y_{6}-y_{7})(y_{1}-y_{4}-y_{6}+y_{7})(y_{1}-y_{4}+y_{6}-y_{7})(y_{1}-y_{4}+y_{6}+y_{7})\\
(y_{1}-y_{4}-y_{5}-y_{8})(y_{1}-y_{4}-y_{5}+y_{8})(y_{1}-y_{4}+y_{5}-y_{8})(y_{1}-y_{4}+y_{5}+y_{8})\\
(y_{1}+y_{4}-y_{6}-y_{7})(y_{1}+y_{4}-y_{6}+y_{7})(y_{1}+y_{4}+y_{6}-y_{7})(y_{1}+y_{4}+y_{6}+y_{7})\\
(y_{1}+y_{4}-y_{5}-y_{8})(y_{1}+y_{4}-y_{5}+y_{8})(y_{1}+y_{4}+y_{5}-y_{8})(y_{1}+y_{4}+y_{5}+y_{8})\\
(y_{1}-y_{3}-y_{6}-y_{8})(y_{1}-y_{3}-y_{6}+y_{8})(y_{1}-y_{3}+y_{6}-y_{8})(y_{1}-y_{3}+y_{6}+y_{8})\\
(y_{1}-y_{3}-y_{5}-y_{7})(y_{1}-y_{3}-y_{5}+y_{7})(y_{1}-y_{3}+y_{5}-y_{7})(y_{1}-y_{3}+y_{5}+y_{7})\\
(y_{1}+y_{3}-y_{6}-y_{8})(y_{1}+y_{3}-y_{6}+y_{8})(y_{1}+y_{3}+y_{6}-y_{8})(y_{1}+y_{3}+y_{6}+y_{8})\\
(y_{1}+y_{3}-y_{5}-y_{7})(y_{1}+y_{3}-y_{5}+y_{7})(y_{1}+y_{3}+y_{5}-y_{7})(y_{1}+y_{3}+y_{5}+y_{7})\\
(y_{1}-y_{2}-y_{7}-y_{8})(y_{1}-y_{2}-y_{7}+y_{8})(y_{1}-y_{2}+y_{7}-y_{8})(y_{1}-y_{2}+y_{7}+y_{8})\\
(y_{1}-y_{2}-y_{5}-y_{6})(y_{1}-y_{2}-y_{5}+y_{6})(y_{1}-y_{2}+y_{5}-y_{6})(y_{1}-y_{2}+y_{5}+y_{6})\\
(y_{1}-y_{2}-y_{3}-y_{4})(y_{1}-y_{2}-y_{3}+y_{4})(y_{1}-y_{2}+y_{3}-y_{4})(y_{1}-y_{2}+y_{3}+y_{4})\\
(y_{1}+y_{2}-y_{7}-y_{8})(y_{1}+y_{2}-y_{7}+y_{8})(y_{1}+y_{2}+y_{7}-y_{8})(y_{1}+y_{2}+y_{7}+y_{8})\\
(y_{1}+y_{2}-y_{5}-y_{6})(y_{1}+y_{2}-y_{5}+y_{6})(y_{1}+y_{2}+y_{5}-y_{6})(y_{1}+y_{2}+y_{5}+y_{6})\\
(y_{1}+y_{2}-y_{3}-y_{4})(y_{1}+y_{2}-y_{3}+y_{4})(y_{1}+y_{2}+y_{3}-y_{4})(y_{1}+y_{2}+y_{3}+y_{4})\Bigr)^{2}
\end{array}\end{equation}

We can restrict this polynomial to $\bigwedge^4\mathcal{H}=\bigwedge^4\CC^8\subset \mathcal{F}_8$. Indeed considering the seven-dimensional Cartan subalgebra spanned by $\ket{E_1},\dots,\ket{E_7}$, the projection of $\text{HDet}_{\Spin(16,\CC)}$ is obtained by dividing by $y_8^{2}$, simplifying, and then setting $y_{8}=0$ in Eq.~\eqref{eq:hdetspin}. One obtains 
\begin{equation}\label{eq:restrictionspin}
 \HDet_{\Spin(16,\CC)|\bigwedge^4\CC^8}(\Psi) = Q^{2}T^{4}
\end{equation}
with
\begin{equation*}
\begin{array}{l}
Q = y_{7}y_{6}y_{5}y_{4}y_{3}y_{2}y_{1}
(y_{3}-y_{4}-y_{5}-y_{6})(y_{3}-y_{4}-y_{5}+y_{6})(y_{3}-y_{4}+y_{5}-y_{6})(y_{3}-y_{4}+y_{5}+y_{6})\\
(y_{3}+y_{4}-y_{5}-y_{6})(y_{3}+y_{4}-y_{5}+y_{6})(y_{3}+y_{4}+y_{5}-y_{6})(y_{3}+y_{4}+y_{5}+y_{6})(y_{2}-y_{4}-y_{5}-y_{7})\\
(y_{2}-y_{4}-y_{5}+y_{7})(y_{2}-y_{4}+y_{5}-y_{7})(y_{2}-y_{4}+y_{5}+y_{7})(y_{2}+y_{4}-y_{5}-y_{7})(y_{2}+y_{4}-y_{5}+y_{7})\\(y_{2}+y_{4}+y_{5}-y_{7})(y_{2}+y_{4}+y_{5}+y_{7})(y_{2}-y_{3}-y_{6}-y_{7})(y_{2}-y_{3}-y_{6}+y_{7})(y_{2}-y_{3}+y_{6}-y_{7})\\
(y_{2}-y_{3}+y_{6}+y_{7})(y_{2}+y_{3}-y_{6}-y_{7})(y_{2}+y_{3}-y_{6}+y_{7})(y_{2}+y_{3}+y_{6}-y_{7})(y_{2}+y_{3}+y_{6}+y_{7})\\
(y_{1}-y_{4}-y_{6}-y_{7})(y_{1}-y_{4}-y_{6}+y_{7})(y_{1}-y_{4}+y_{6}-y_{7})(y_{1}-y_{4}+y_{6}+y_{7})(y_{1}+y_{4}-y_{6}-y_{7})\\
(y_{1}+y_{4}-y_{6}+y_{7})(y_{1}+y_{4}+y_{6}-y_{7})(y_{1}+y_{4}+y_{6}+y_{7})(y_{1}-y_{3}-y_{5}-y_{7})(y_{1}-y_{3}-y_{5}+y_{7})\\
(y_{1}-y_{3}+y_{5}-y_{7})(y_{1}-y_{3}+y_{5}+y_{7})(y_{1}+y_{3}-y_{5}-y_{7})(y_{1}+y_{3}-y_{5}+y_{7})(y_{1}+y_{3}+y_{5}-y_{7})\\(y_{1}+y_{3}+y_{5}+y_{7})(y_{1}-y_{2}-y_{5}-y_{6})(y_{1}-y_{2}-y_{5}+y_{6})(y_{1}-y_{2}+y_{5}-y_{6})(y_{1}-y_{2}+y_{5}+y_{6})\\
(y_{1}-y_{2}-y_{3}-y_{4})(y_{1}-y_{2}-y_{3}+y_{4})(y_{1}-y_{2}+y_{3}-y_{4})(y_{1}-y_{2}+y_{3}+y_{4})(y_{1}+y_{2}-y_{5}-y_{6})\\
(y_{1}+y_{2}-y_{5}+y_{6})(y_{1}+y_{2}+y_{5}-y_{6})(y_{1}+y_{2}+y_{5}+y_{6})(y_{1}+y_{2}-y_{3}-y_{4})(y_{1}+y_{2}-y_{3}+y_{4})\\
(y_{1}+y_{2}+y_{3}-y_{4})(y_{1}+y_{2}+y_{3}+y_{4}),
\end{array}
\end{equation*}
and
\begin{equation*}
\begin{array}{l}
T = (y_{5}-y_{6}-y_{7})(y_{5}-y_{6}+y_{7})(y_{5}+y_{6}-y_{7})(y_{5}+y_{6}+y_{7})(y_{3}-y_{4}-y_{7})(y_{3}-y_{4}+y_{7})\\
(y_{3}+y_{4}-y_{7})(y_{3}+y_{4}+y_{7})(y_{2}-y_{4}-y_{6})(y_{2}-y_{4}+y_{6})(y_{2}+y_{4}-y_{6})(y_{2}+y_{4}+y_{6})\\
(y_{2}-y_{3}-y_{5})(y_{2}-y_{3}+y_{5})(y_{2}+y_{3}-y_{5})(y_{2}+y_{3}+y_{5})(y_{1}-y_{4}-y_{5})(y_{1}-y_{4}+y_{5})\\
(y_{1}+y_{4}-y_{5})(y_{1}+y_{4}+y_{5})(y_{1}-y_{3}-y_{6})(y_{1}-y_{3}+y_{6})(y_{1}+y_{3}-y_{6})(y_{1}+y_{3}+y_{6})\\
(y_{1}-y_{2}-y_{7})(y_{1}-y_{2}+y_{7})(y_{1}+y_{2}-y_{7})(y_{1}+y_{2}+y_{7}).
 \end{array}
\end{equation*}

\vspace{.5em}
The factor $Q$ of degree 63 in the expression of $\HDet_{\Spin(16,\CC)|\bigwedge^4\CC^8}$ in Eq.~\eqref{eq:restrictionspin} corresponds to the expression $\Delta_{E_7|\bigwedge^4\CC^8}$ as computed in \cite{HolweckOeding} from the grading $\mathfrak{e}_7=\mathfrak{s}\mathfrak{l}_8\oplus\bigwedge^4 \CC^8$ \cite{antonyan1981,oeding2022}. This last polynomial is the restriction to semi-simple elements of the dual equation of the Grassmannian of four-planes in $\CC^8$. In the quantum information literature it would be the analogue of Cayley's hyperdeterminant for $4$ fermions with  $8$-single particles states and as such is interesting to study fermionic entanglement. In \cite{HolweckOeding} we also showed how this equation projects to the hyperdeterminant of format $2\times 2\times 2\times 2$. The second factor, which has degree $28$ and is named $T$ in Eq.~\eqref{eq:restrictionspin}, is by construction another $\text{SL}_8(\CC)$ invariant polynomial for the module $\bigwedge^4 \CC^8$.

\section{A combinatorial interpretation}\label{sec:fano}
\begin{figure}[!h]
 \centering
 \scalebox{0.6}
 {
 \begin{tikzpicture}[scale=4]
	 \tikzstyle{vertex}=[circle,minimum size=20pt,inner sep=0pt]
	 \tikzstyle{selected vertex} = [vertex, fill=red!]
	 \tikzstyle{selected edge} = [draw,dashed,thick,red!]
	 \tikzstyle{selected edge2} = [draw,dashed,thick,green!]
	 \tikzstyle{selected edge3} = [draw,dashed,thick,blue!]
	 \tikzstyle{edge} = [draw,thick,-,black]
	 \node[vertex] (v0) at (0,0) {$y_1$};
	 \node[vertex] (v1) at (0,1) {$y_3$};
	 \node[vertex] (v2) at (1,0) {$y_2$};
	 \node[vertex] (v3) at (1,1) {$y_4$};
	 \node[vertex] (v4) at (0.23, 0.4) {$y_5$};
	 \node[vertex] (v5) at (0.23,1.4) {$y_7$};
	 \node[vertex] (v6) at (1.23,0.4) {$y_6$};
	 \node[vertex] (v7) at (1.23,1.4) {$y_8$};
	 \draw[edge] (v0) -- (v1) -- (v3) -- (v2) -- (v0);
	 \draw[edge] (v0) -- (v4) -- (v5) -- (v1) -- (v0);
	 \draw[edge] (v2) -- (v6) -- (v7) -- (v3) -- (v2);
	 \draw[edge] (v4) -- (v6) -- (v7) -- (v5) -- (v4);
	 \draw[edge] (v0) -- (v2);
	 \draw[edge] (v2) -- (v6);
	 \draw[edge] (v6) -- (v4);
	 \draw[edge] (v4) -- (v5);	
	 \draw[selected edge] (v0) -- (v2);
     \draw[selected edge] (v2) -- (v3);
     \draw[selected edge] (v3) -- (v1);
     \draw[selected edge] (v1) -- (v0);
     
     \node[vertex] (vv0) at (1.5,0) {$y_1$};
	 \node[vertex] (vv1) at (1.5,1) {$y_3$};
	 \node[vertex] (vv2) at (2.5,0) {$y_2$};
	 \node[vertex] (vv3) at (2.5,1) {$y_4$};
	 \node[vertex] (vv4) at (1.73, 0.4) {$y_5$};
	 \node[vertex] (vv5) at (1.73,1.4) {$y_7$};
	 \node[vertex] (vv6) at (2.73,0.4) {$y_6$};
	 \node[vertex] (vv7) at (2.73,1.4) {$y_8$};
	 \draw[edge] (vv0) -- (vv1) -- (vv3) -- (vv2) -- (vv0);
	 \draw[edge] (vv0) -- (vv4) -- (vv5) -- (vv1) -- (vv0);
	 \draw[edge] (vv2) -- (vv6) -- (vv7) -- (vv3) -- (vv2);
	 \draw[edge] (vv4) -- (vv6) -- (vv7) -- (vv5) -- (vv4);
	 \draw[edge] (vv0) -- (vv2);
	 \draw[edge] (vv2) -- (vv6);
	 \draw[edge] (vv6) -- (vv4);
	 \draw[edge] (vv4) -- (vv5);	
	 \draw[selected edge2] (vv0) -- (vv4);
	 \draw[selected edge2] (vv3) -- (vv7);
	 \draw[selected edge2] (vv0) -- (vv3);
	 \draw[selected edge2] (vv4) -- (vv7);
	 
	 \node[vertex] (vvv0) at (3,0) {$y_1$};
	 \node[vertex] (vvv1) at (3,1) {$y_3$};
	 \node[vertex] (vvv2) at (4,0) {$y_2$};
	 \node[vertex] (vvv3) at (4,1) {$y_4$};
	 \node[vertex] (vvv4) at (3.23, 0.4) {$y_5$};
	 \node[vertex] (vvv5) at (3.23,1.4) {$y_7$};
	 \node[vertex] (vvv6) at (4.23,0.4) {$y_6$};
	 \node[vertex] (vvv7) at (4.23,1.4) {$y_8$};
	 \draw[edge] (vvv0) -- (vvv1) -- (vvv3) -- (vvv2) -- (vvv0);
	 \draw[edge] (vvv0) -- (vvv4) -- (vvv5) -- (vvv1) -- (vvv0);
	 \draw[edge] (vvv2) -- (vvv6) -- (vvv7) -- (vvv3) -- (vvv2);
	 \draw[edge] (vvv4) -- (vvv6) -- (vvv7) -- (vvv5) -- (vvv4);
	 \draw[edge] (vvv0) -- (vvv2);
	 \draw[edge] (vvv2) -- (vvv6);
	 \draw[edge] (vvv6) -- (vvv4);
	 \draw[edge] (vvv4) -- (vvv5);	
	 \draw[selected edge3] (vvv0) -- (vvv6);
	 \draw[selected edge3] (vvv0) -- (vvv3);
	 \draw[selected edge3] (vvv0) -- (vvv5);
	 \draw[selected edge3] (vvv3) -- (vvv6);
	 \draw[selected edge3] (vvv3) -- (vvv5);
	 \draw[selected edge3] (vvv6) -- (vvv5);
 \end{tikzpicture}
 }
 \caption{Planes of the cube in $(\ZZ_2)^3$: $6$ planes as the ``regular'' faces, $6$ planes through the ``opposite edges'' and $2$ ``tetrahedron''-planes.}\label{fig:plane}
\end{figure}
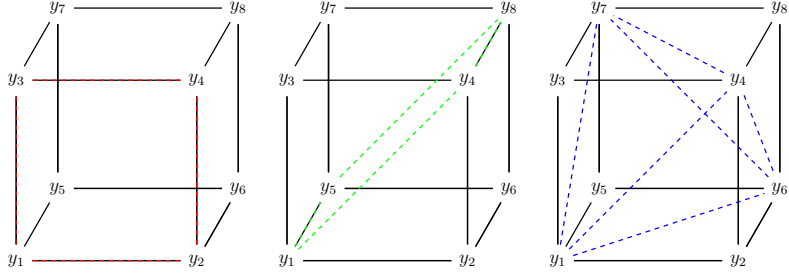

The invariant polynomials $\text{HDet}_{\Spin(16,\CC)}$ and $\text{HDet}_{\Spin(16,\CC)|\bigwedge^4\CC^8}$ described on semi-simple elements have nice combinatorial interpretations. Consider the cube in $(\ZZ_2)^3$ with vertices 
\begin{equation}
 \begin{array}{cccc}
 y_1=(0,0,0), & y_2=(0,1,0), & y_3=(1,0,0), & y_4=(1,1,0),\\
 y_5=(0,0,1), & y_6=(0,1,1), & y_7=(1,0,1), & y_8=(1,1,1).
 \end{array}
\end{equation}
The cube in $(\ZZ_2)^3$ comprises $8$ points, $28$ lines and $14$ planes. The planes are represented in Figure~\ref{fig:plane}. 
They are planes in the sense that their defining equation is linear. For instance the tetrahedron-plane of Figure~\ref{fig:plane} is given in the $(z_1,z_2,z_3)$ coordinates of $(\ZZ_2)^3$ by the equation $z_1+z_2+z_3=0$.
The $14$ planes of the cube in $\ZZ_3$ generate the $8\times 14=112$ linear forms of Eq.~\eqref{eq:hdetspin} that involve $4$ variables. Including the linear forms $y_1,\dots, y_8$ one gets the product of $120$ linear forms (every linear form being squared to make a degree $240$ polynomial). Thus we have the following formula parametrized by vertices and planes of the cube in $(\ZZ_3)^2$:
\begin{equation}
 \text{HDet}_{\Spin(16,\CC)}(\Psi)=\Bigl (\prod_{y_{i}\text{-vertices}}y_{i}\prod_{(y_{i_1},y_{i_2},y_{i_3}, y_{i_4})\text{-planes}} (y_{i_1}\pm y_{i_2}\pm y_{i_3}\pm y_{i_4})\Bigr)^2,
\end{equation}
with the convention that $i_1<i_2<i_3<i_4$.

The restriction of $\text{HDet}_{\Spin(16,\CC)}$ to $\bigwedge^4\CC^8$ was obtained by eliminating $y_8$ in Eq.~\eqref{eq:hdetspin}. In our combinatorial  picture it could be obtained by projecting the planes of $(\ZZ_2)^3$ to the projective space $PG(2,2)$ where the coordinate associated to $y_8$ is considered as the center. The $2$-dimensional projective space over $\ZZ_2$ is well-known as the Fano-plane. The planes passing through the center will project to lines while the planes that do not pass though the center will correspond to affine planes in $PG(2,2)$.
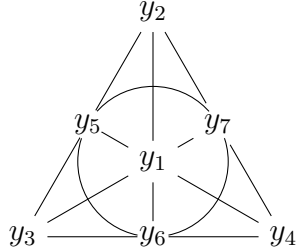
\begin{figure}[!h]
\centering
 \scalebox{1}
 {\begin{tikzpicture}
\tikzstyle{point}=[ball color=white, circle, draw=white, inner sep=0.2cm]
\node (v7) at (0,0)  {$y_1$};
\draw (0,0) circle (1cm);
\node (v1) at (90:2cm)  {$y_2$};
\node (v2) at (210:2cm)  {$y_3$};
\node (v4) at (330:2cm) {$y_4$};
\node (v3) at (150:1cm)  {$ $};
\node (v6) at (270:1cm) {$ $};
\node (v5) at (30:1cm)  {$y_7$};

\draw (v1) -- (v3) -- (v2);
\draw (v2) -- (v6) -- (v4);
\draw (v4) -- (v5) -- (v1);
\draw (v3) -- (v7) -- (v4);
\draw (v5) -- (v7) -- (v2);
\draw (v6) -- (v7) -- (v1);
\draw[fill=white,color=white] (v3) circle[radius=0.2];
\node (v13) at (150:1cm) {$y_5$};
\draw[fill=white,color=white] (v6) circle[radius=0.2];
\node (v16) at (v6) {$y_6$};
\draw[fill=white,color=white] (v3) circle[radius=0.2];
\node (v13) at (150:1cm) {$y_5$};
\draw[fill=white,color=white] (v5) circle[radius=0.2];
\node (v15) at (v5) {$y_7$};
\end{tikzpicture}}
\caption{The Fano plane obtained by considering the projective space associated to the cube $(\ZZ_2)^3$ when the point corresponding to variable $y_8$ is chosen as the center. The planes of $(\ZZ_2)^3$ passing through $y_8$ are sent to lines and the planes that do not go through $y_8$ are mapped to affine plane of the Fano plane.} 
\end{figure}\label{fig:fanoplane}

Therefore one can see Eq.~\eqref{eq:restrictionspin} as a product of linear forms parametrized by the vertices, affine planes, and lines in the Fano plane:
\begin{multline}
 \text{HDet}_{\Spin(16,\CC)|\bigwedge^4 \CC^8}(\Psi)=\\
\Bigl(\prod_{y_{i}\text{-vertices}}y_{i} \prod_{(y_{i_1},y_{i_2},y_{i_3}, y_{i_4})\text{-planes}}(y_{i_1}\pm y_{i_2}\pm y_{i_3}\pm y_{i_4})\Bigr)^2 \Bigl (\prod_{(y_{j_1},y_{j_2},y_{j_3})\text{-lines}} (y_{i_1}\pm y_{i_2}\pm y_{i_3})\Bigr)^4,
\end{multline}
with the convention that $i_1<i_2<i_3<i_4$ and $j_1<j_2<j_3$.
This observation reveals an interesting connection between the Fano plane and the dual of the Grassmannian $\text{Gr}(4,8)$. 

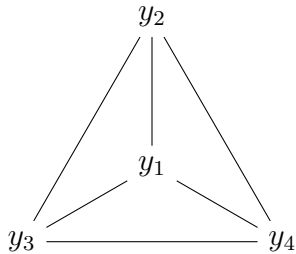
\begin{figure}[!h]
\centering
 \scalebox{1}
 {\begin{tikzpicture}
\tikzstyle{point}=[ball color=white, circle, draw=white, inner sep=0.2cm]
\node (v1) at (0,0)  {$y_1$};
\node (v2) at (90:2cm)  {$y_2$};
\node (v3) at (210:2cm)  {$y_3$};
\node (v4) at (330:2cm) {$y_4$};

\draw (v3) -- (v2);
\draw (v2)  -- (v4);
\draw (v4)  -- (v1);
\draw (v3) --  (v4);
\draw (v3) -- (v1);
\draw (v1) -- (v2);
\end{tikzpicture}}
\caption{The affine plane obtained by removing the projective line $y_5y_6y_7$. } \label{fig:plane567}
\end{figure}

Furthermore one can eliminate three more variables and restrict to a four-dimensional Cartan algebra. Suppose we remove the line $y_5y_6y_7$ of Figure (\ref{fig:fanoplane}) and we keep the collinearity relations. One gets the affine plane depicted in Figure~\ref{fig:plane567}.
Considering the monomials given by the lines of this affine plane one gets the polynomial,
\begin{equation}
 \Delta(y_1,y_2,y_3,y_4)=\prod_{(y_i,y_j)\text{-lines}, i<j} (y_i\pm y_j)^2.
\end{equation}

This last equation is nothing but the expression of the $2\times 2\times 2\times 2$ hyperdeterminant (see \cite{luque2003}) restricted to the semisimple elements of Eq.~\eqref{eq:semisimple} for $y_5=y_6=y_7=y_8=0$. For an understanding of the four-qubit entanglement classification based on the study of $\Delta$ one refers to \cite{holweck_4qubit2} and for other algebraic geometry techniques for this classification see \cite{masoud20}.

 \section{Conclusion}\label{sec:conclusion}
In this paper we evaluate for a generic element of the fermionic Fock space, the analogue of the Cayley's hyperdeterminant in the context of the spinor representation of $\Spin(16,\CC)$. This work was motivated by considerations from quantum information theory as the fermionic Fock space and its spinor representation is a framework to describe many different multipartite Hilbert spaces and their corresponding SLOCC actions. In this setting algebraic invariants are interesting to compute, either to distinguish different classes of entanglement by separating the classes when the polynomials vanish or not, or by measuring some amount of entanglement when we evaluate those invariants.

It is interesting to try to express a given invariant in terms of the fundamental invariants of the corresponding representation. It is known \cite{LevayHolweck} that the ring of invariants $\CC[\mathcal{F}^\pm]^{\Spin(16,\CC)}$ is generated by $8$ polynomials of degree $2,8,12,14,18,20,24$ and $30$. Those polynomials can be evaluated on a generic fermionic Fock states by the same techniques as we used in this paper (See the Appendix). However an expression of $\text{HDet}_{\Spin(16,\CC)}$ in terms of those invariants sounds quite difficult to obtain due to the number of monomials involved for trying a naive interpolation.
\section*{Acknowledgements}
This work was partially done during the visit of FH to Auburn University during the academic year 2021-2022. FH
 thanks Pr. Ash Abebe and his staff of the AU Mathematics and Statistics department for making his visiting year in Auburn possible. FH was partially supported by the Conseil
Régional de Bourgogne Franche-Comté (GeoQuant project). Both authors acknowledge partial support from the Thomas Jefferson foundation. 
\begin{appendix}

\section*{Appendix: Fundamental invariants on fermionic Fock space with $N=8$}\label{sec:appendix}
One evaluates the fundamental invariants of $\CC[\mathcal{F}^{\pm}_8]^{\text{Spin}(16)}$ on the generic semi-simple element given by Eq.~\eqref{eq:semisimple}. To do so, we restrict to the Cartan of the spin module, the fundamental invariants of $\CC[\mathfrak{e}_8]^{E_8}$. Those invariants can be obtained by the Chevalley restriction Theorem (see Section~\ref{sec:e8}) and the observation that the power sum expression 
\[f_d=\sum_{\alpha \in R} \alpha^d, \text{ with } R \text{ is the set of roots of } E_8 \]
is invariant under the action of the the Weyl group $E_8$. A set of fundamental invariants of $\CC[\mathcal{F}^{\pm}_8]^{\text{Spin}(16)}$ consists of homogeneous polynomials of degrees $d=2,8,12,14,18,20,24,30$ (one of each). One may obtain a set of invariants  by first calculating the expression of the polynomials $f_k$ for $k=2,8,12,14,18,20,24,30$ in terms of the $x_i$'s (See Eq.~\eqref{eq:Roots}) and then expressing those polynomials in terms of the $y_i$'s, i.e. restricting them to a Cartan subalgebra of $\mathcal{F}^{\pm}_8$ (see Eq.~\eqref{eq:yx}). It turns out that the invariant expressions obtained in this way are non-trivial and generate the invariant ring.
Moreover, the fundamental invariants restricted to the $y_{i}$'s have an additional $\mathfrak{S}_{8}$-symmetry. Respectively the invariants have 1, 7, 14, 17, 29, 38, 57, and 93 terms up to symmetry that we list below. 

\vspace{1em}
\noindent $\begin{smallmatrix}
f_{2}:& y_{1}^{2}\\
\end{smallmatrix}
$
\vspace{1em}

\noindent $\begin{smallmatrix}
f_{8}:& 6\,061\,y_{1}^{8},\:7\,196\,y_{1}^{6}y_{2}^{2},\:17\,990\,y_{1}^{4}y_{2}^{4},\:35\,980\,y_{1}^{4}y_{2}^{2}y_{3}^{2}, 
 \:-28\,560\,y_{1}^{5}y_{2}y_{3}y_{4},\:-95\,200\, y_{1}^{3}y_{2}^{3}y_{3}y_{4},\:215\,880\,y_{1}^{2}y_{2}^{2}y_{3}^{2}y_{4}^{2}\hfill 
\end{smallmatrix}$
\vspace{1em}

\noindent $\begin{smallmatrix}
f_{12}:
&
1\,407\,661\,y_{1}^{12},\:270\,402\,y_{1}^{10}y_{2}^{2},\:2\,028\,015\,y_{1}^{8}y_{2}^{4},\:3\,785\,628\,y_{1}^{6}y_{2}^{6}, \:4\,056\,030\,y_{1}^{8}y_{2}^{2}y_{3}^{2}, \:18\,928\,140\,y_{1}^{6}y_{2}^{4}y_{3}^{2},\:47\,320\,350\,y_{1}^{4}y_{2}^{4}y_{3}^{4},\hfill \\ &   
 \:-1\,801\,800\,y_{1}^{9}y_{2}y_{3}y_{4}, \:-21\,621\,600\,y_{1}^{7}y_{2}^{3}y_{3}y_{4},  \:-45\,405\,360\,y_{1}^{5}y_{2}^{5}y_{3}y_{4},\:-151\,351\,200\,y_{1}^{5}y_{2}^{3}y_{3}^{3}y_{4},\:113\,568\,840\,y_{1}^{6}y_{2}^{2}y_{3}^{2}y_{4}^{2}, \hfill \\ &   
  \:283\,922\,100\,y_{1}^{4}y_{2}^{4}y_{3}^{2}y_{4}^{2},\:-504\,504\,000\,y_{1}^{3}y_{2}^{3}y_{3}^{3}y_{4}^{3} \hfill
\end{smallmatrix}$
\vspace{1em}

\noindent $\begin{smallmatrix}
f_{14}:& 1\,723\,681\,y_{1}^{14},\:114\,695\,y_{1}^{12}y_{2}^{2},\:1\,261\,645\,y_{1}^{10}y_{2}^{4},\:3\,784\,935\,y_{1}^{8}y_{2}^{6}, \:2\,523\,290\,y_{1}^{10}y_{2}^{2}y_{3}^{2},\:18\,924\,675\,y_{1}^{8}y_{2}^{4}y_{3}^{2},\:35\,326\,060\,y_{1}^{6}y_{2}^{6}y_{3}^{2}, \hfill \\ &  
\:88\,315\,150\,y_{1}^{6}y_{2}^{4}y_{3}^{4},\:-917\,448\,y_{1}^{11}y_{2}y_{3}y_{4},
\:-16\,819\,880\,y_{1}^{9}y_{2}^{3}y_{3}y_{4},\:-60\,551\,568\,y_{1}^{7}y_{2}^{5}y_{3}y_{4},\:-201\,838\,560\,y_{1}^{7}y_{2}^{3}y_{3}^{3}y_{4},\hfill \\ &  
\:-423\,860\,976\,y_{1}^{5}y_{2}^{5}y_{3}^{3}y_{4},\:113\,548\,050\,y_{1}^{8}y_{2}^{2}y_{3}^{2}y_{4}^{2},\:529\,890\,900\,y_{1}^{6}y_{2}^{4}y_{3}^{2}y_{4}^{2},
\:1\,324\,727\,250\,y_{1}^{4}y_{2}^{4}y_{3}^{4}y_{4}^{2},\:-1\,412\,869\,920\,y_{1}^{5}y_{2}^{3}y_{3}^{3}y_{4}^{3}\hfill 
\end{smallmatrix}$
\vspace{1em}

\noindent $\begin{smallmatrix}
f_{18}:& 5\,727\,234\,733\,y_{1}^{18},\:40\,108\,185\,y_{1}^{16}y_{2}^{2},\:802\,163\,700\,y_{1}^{14}y_{2}^{4},\:4\,866\,459\,780\,y_{1}^{12}y_{2}^{6}, \:11\,470\,940\,910\,y_{1}^{10}y_{2}^{8},\:1\,604\,327\,400\,y_{1}^{14}y_{2}^{2}y_{3}^{2},\hfill \\ &
\:24\,332\,298\,900\,y_{1}^{12}y_{2}^{4}y_{3}^{2},\:107\,062\,115\,160\,y_{1}^{10}y_{2}^{6}y_{3}^{2},\:172\,064\,113\,650\,y_{1}^{8}y_{2}^{8}y_{3}^{2},\:267\,655\,287\,900\,y_{1}^{10}y_{2}^{4}y_{3}^{4},\:802\,965\,863\,700\,y_{1}^{8}y_{2}^{6}y_{3}^{4}, \hfill \\ &
\:1\,498\,869\,612\,240\,y_{1}^{6}y_{2}^{6}y_{3}^{6},\:-427\,817\,376\,y_{1}^{15}y_{2}y_{3}y_{4},\:-14\,973\,608\,160\,y_{1}^{13}y_{2}^{3}y_{3}y_{4},\:-116\,794\,143\,648\,y_{1}^{11}y_{2}^{5}y_{3}y_{4}, \hfill \\ &
\:-305\,889\,423\,840\,y_{1}^{9}y_{2}^{7}y_{3}y_{4},\:-389\,313\,812\,160\,y_{1}^{11}y_{2}^{3}y_{3}^{3}y_{4},\: -2\,141\,225\,966\,880\,y_{1}^{9}y_{2}^{5}y_{3}^{3}y_{4},\:-3\,670\,673\,086\,080\,y_{1}^{7}y_{2}^{7}y_{3}^{3}y_{4}, \hfill \\ &
\:-7\,708\,413\,480\,768\,y_{1}^{7}y_{2}^{5}y_{3}^{5}y_{4},\:145\,993\,793\,400\,y_{1}^{12}y_{2}^{2}y_{3}^{2}y_{4}^{2},\:1\,605\,931\,727\,400\,y_{1}^{10}y_{2}^{4}y_{3}^{2}y_{4}^{2},\:4\,817\,795\,182\,200\,y_{1}^{8}y_{2}^{6}y_{3}^{2}y_{4}^{2},\hfill \\ &
\:12\,044\,487\,955\,500\,y_{1}^{8}y_{2}^{4}y_{3}^{4}y_{4}^{2},\:22\,483\,044\,183\,600\,y_{1}^{6}y_{2}^{6}y_{3}^{4}y_{4}^{2},\: -7\,137\,419\,889\,600\,y_{1}^{9}y_{2}^{3}y_{3}^{3}y_{4}^{3}, \:-25\,694\,711\,602\,560\,y_{1}^{7}y_{2}^{5}y_{3}^{3}y_{4}^{3},\hfill \\ &
\:-53\,958\,894\,365\,376\,y_{1}^{5}y_{2}^{5}y_{3}^{5}y_{4}^{3},
\:56\,207\,610\,459\,000\,y_{1}^{6}y_{2}^{4}y_{3}^{4}y_{4}^{4} \hfill
\end{smallmatrix}
$
\vspace{1em}

\noindent $\begin{smallmatrix}
f_{20}:& 91\,628\,415\,661\,y_{1}^{20},\:199\,229\,630\,y_{1}^{18}y_{2}^{2},\:5\,080\,355\,565\,y_{1}^{16}y_{2}^{4},\:40\,642\,844\,520\,y_{1}^{14}y_{2}^{6},\:132\,089\,244\,690\,y_{1}^{12}y_{2}^{8},
\hfill \\ &
193\,730\,892\,212\,y_{1}^{10}y_{2}^{10}, \:10\,160\,711\,130\,y_{1}^{16}y_{2}^{2}y_{3}^{2},\:203\,214\,222\,600\,y_{1}^{14}y_{2}^{4}y_{3}^{2},\:1\,232\,832\,950\,440\,y_{1}^{12}y_{2}^{6}y_{3}^{2},
\hfill \\ &
2\,905\,963\,383\,180\,y_{1}^{10}y_{2}^{8}y_{3}^{2}, \:3\,082\,082\,376\,100\,y_{1}^{12}y_{2}^{4}y_{3}^{4},\:13\,561\,162\,454\,840\,y_{1}^{10}y_{2}^{6}y_{3}^{4},\:21\,794\,725\,373\,850\,y_{1}^{8}y_{2}^{8}y_{3}^{4},
\hfill \\ &
40\,683\,487\,364\,520\,y_{1}^{8}y_{2}^{6}y_{3}^{6}, \:-2\,390\,751\,000\,y_{1}^{17}y_{2}y_{3}y_{4},\:-108\,380\,712\,000\,y_{1}^{15}y_{2}^{3}y_{3}y_{4},
\:-1\,137\,997\,476\,000\,y_{1}^{13}y_{2}^{5}y_{3}y_{4},\hfill \\ &
\:-4\,226\,847\,768\,000\,y_{1}^{11}y_{2}^{7}y_{3}y_{4}, \:-6\,457\,684\,090\,000\,y_{1}^{9}y_{2}^{9}y_{3}y_{4},
\:-3\,793\,324\,920\,000\,y_{1}^{13}y_{2}^{3}y_{3}^{3}y_{4},\hfill \\ &
\:-29\,587\,934\,376\,000\,y_{1}^{11}y_{2}^{5}y_{3}^{3}y_{4},\:-77\,492\,209\,080\,000\,y_{1}^{9}y_{2}^{7}y_{3}^{3}y_{4}, 
\:-162\,733\,639\,068\,000\,y_{1}^{9}y_{2}^{5}y_{3}^{5}y_{4},\hfill \\ &
\:-278\,971\,952\,688\,000\,y_{1}^{7}y_{2}^{7}y_{3}^{5}y_{4},\:1\,219\,285\,335\,600\,y_{1}^{14}y_{2}^{2}y_{3}^{2}y_{4}^{2},
\:18\,492\,494\,256\,600\,y_{1}^{12}y_{2}^{4}y_{3}^{2}y_{4}^{2}, \hfill \\ &
\:81\,366\,974\,729\,040\,y_{1}^{10}y_{2}^{6}y_{3}^{2}y_{4}^{2},\:130\,768\,352\,243\,100\,y_{1}^{8}y_{2}^{8}y_{3}^{2}y_{4}^{2},
\:203\,417\,436\,822\,600\,y_{1}^{10}y_{2}^{4}y_{3}^{4}y_{4}^{2},\hfill \\ &
\:610\,252\,310\,467\,800\,y_{1}^{8}y_{2}^{6}y_{3}^{4}y_{4}^{2}, \:1\,139\,137\,646\,206\,560\,y_{1}^{6}y_{2}^{6}y_{3}^{6}y_{4}^{2},
\:-98\,626\,447\,920\,000\,y_{1}^{11}y_{2}^{3}y_{3}^{3}y_{4}^{3},\hfill \\ &
\:-542\,445\,463\,560\,000\,y_{1}^{9}y_{2}^{5}y_{3}^{3}y_{4}^{3},\:-929\,906\,508\,960\,000\,y_{1}^{7}y_{2}^{7}y_{3}^{3}y_{4}^{3},
\:-1\,952\,803\,668\,816\,000\,y_{1}^{7}y_{2}^{5}y_{3}^{5}y_{4}^{3},\hfill \\ &
\:1\,525\,630\,776\,169\,500\,y_{1}^{8}y_{2}^{4}y_{3}^{4}y_{4}^{4},\:2\,847\,844\,115\,516\,400\,y_{1}^{6}y_{2}^{6}y_{3}^{4}y_{4}^{4},
\:-4\,100\,887\,704\,513\,600\,y_{1}^{5}y_{2}^{5}y_{3}^{5}y_{4}^{5} \hfill
\end{smallmatrix}
$

\medskip
\noindent $\begin{smallmatrix}
f_{24}:& 23\,456\,287\,206\,061\,y_{1}^{24},\:4\,630\,511\,892\,y_{1}^{22}y_{2}^{2},\:178\,274\,707\,842\,y_{1}^{20}y_{2}^{4},\:2\,258\,146\,299\,332\,y_{1}^{18}y_{2}^{6},
\:12\,339\,156\,564\,207\,y_{1}^{16}y_{2}^{8},
\hfill \\ &
\:32\,904\,417\,504\,552\,y_{1}^{14}y_{2}^{10},\:45\,368\,212\,013\,852\,y_{1}^{12}y_{2}^{12},\:356\,549\,415\,684\,y_{1}^{20}y_{2}^{2}y_{3}^{2},\:11\,290\,731\,496\,660\,y_{1}^{18}y_{2}^{4}y_{3}^{2},\hfill \\ & 
\:115\,165\,461\,265\,932\,y_{1}^{16}y_{2}^{6}y_{3}^{2},\:493\,566\,262\,568\,280\,y_{1}^{14}y_{2}^{8}y_{3}^{2},\:998\,100\,664\,304\,744\,y_{1}^{12}y_{2}^{10}y_{3}^{2},\:287\,913\,653\,164\,830\,y_{1}^{16}y_{2}^{4}y_{3}^{4},\hfill \\ & 
\:2\,303\,309\,225\,318\,640\,y_{1}^{14}y_{2}^{6}y_{3}^{4},\:7\,485\,754\,982\,285\,580\,y_{1}^{12}y_{2}^{8}y_{3}^{4},\:10\,979\,107\,307\,352\,184\,y_{1}^{10}y_{2}^{10}y_{3}^{4},
\hfill \\ & 
\:13\,973\,409\,300\,266\,416\,y_{1}^{12}y_{2}^{6}y_{3}^{6},
\:32\,937\,321\,922\,056\,552\,y_{1}^{10}y_{2}^{8}y_{3}^{6},\:52\,934\,981\,660\,448\,030\,y_{1}^{8}y_{2}^{8}y_{3}^{8},\hfill \\ & 
\:-67\,914\,166\,320\,y_{1}^{21}y_{2}y_{3}y_{4},\:-4\,753\,991\,642\,400\,y_{1}^{19}y_{2}^{3}y_{3}y_{4},\:-81\,293\,257\,085\,040\,y_{1}^{17}y_{2}^{5}y_{3}y_{4},\hfill \\ & 
\:-526\,470\,617\,312\,640\,y_{1}^{15}y_{2}^{7}y_{3}y_{4},\:-1\,535\,539\,300\,495\,200\,y_{1}^{13}y_{2}^{9}y_{3}y_{4},\:-2\,177\,673\,917\,065\,920\,y_{1}^{11}y_{2}^{11}y_{3}y_{4},\hfill \\ & 
\:-270\,977\,523\,616\,800\,y_{1}^{17}y_{2}^{3}y_{3}^{3}y_{4},\:-3\,685\,294\,321\,188\,480\,y_{1}^{15}y_{2}^{5}y_{3}^{3}y_{4},\:-18\,426\,471\,605\,942\,400\,y_{1}^{13}y_{2}^{7}y_{3}^{3}y_{4},\hfill \\ & 
\:-39\,924\,021\,812\,875\,200\,y_{1}^{11}y_{2}^{9}y_{3}^{3}y_{4},\:-38\,695\,590\,372\,479\,040\,y_{1}^{13}y_{2}^{5}y_{3}^{5}y_{4},\:-143\,726\,478\,526\,350\,720\,y_{1}^{11}y_{2}^{7}y_{3}^{5}y_{4},\hfill \\ & 
\:-219\,582\,119\,970\,813\,600\,y_{1}^{9}y_{2}^{9}y_{3}^{5}y_{4},\:-376\,426\,491\,378\,537\,600\,y_{1}^{9}y_{2}^{7}y_{3}^{7}y_{4},\:67\,744\,388\,979\,960\,y_{1}^{18}y_{2}^{2}y_{3}^{2}y_{4}^{2},\hfill \\ & 
\:1\,727\,481\,918\,988\,980\,y_{1}^{16}y_{2}^{4}y_{3}^{2}y_{4}^{2},\:13\,819\,855\,351\,911\,840\,y_{1}^{14}y_{2}^{6}y_{3}^{2}y_{4}^{2},\:44\,914\,529\,893\,713\,480\,y_{1}^{12}y_{2}^{8}y_{3}^{2}y_{4}^{2},\hfill \\ & 
\:65\,874\,643\,844\,113\,104\,y_{1}^{10}y_{2}^{10}y_{3}^{2}y_{4}^{2},\:34\,549\,638\,379\,779\,600\,y_{1}^{14}y_{2}^{4}y_{3}^{4}y_{4}^{2},\:209\,601\,139\,503\,996\,240\,y_{1}^{12}y_{2}^{6}y_{3}^{4}y_{4}^{2},\hfill \\ & 
\:494\,059\,828\,830\,848\,280\,y_{1}^{10}y_{2}^{8}y_{3}^{4}y_{4}^{2},\:922\,245\,013\,817\,583\,456\,y_{1}^{10}y_{2}^{6}y_{3}^{6}y_{4}^{2},\:1\,482\,179\,486\,492\,544\,840\,y_{1}^{8}y_{2}^{8}y_{3}^{6}y_{4}^{2},\hfill \\ & 
\:-12\,284\,314\,403\,961\,600\,y_{1}^{15}y_{2}^{3}y_{3}^{3}y_{4}^{3},\:-128\,985\,301\,241\,596\,800\,y_{1}^{13}y_{2}^{5}y_{3}^{3}y_{4}^{3},\:-479\,088\,261\,754\,502\,400\,y_{1}^{11}y_{2}^{7}y_{3}^{3}y_{4}^{3},\hfill \\ & 
\:-731\,940\,399\,902\,712\,000\,y_{1}^{9}y_{2}^{9}y_{3}^{3}y_{4}^{3},\:-1\,006\,085\,349\,684\,455\,040\,y_{1}^{11}y_{2}^{5}y_{3}^{5}y_{4}^{3},\:-2\,634\,985\,439\,649\,763\,200\,y_{1}^{9}y_{2}^{7}y_{3}^{5}y_{4}^{3},\hfill \\ & 
\:-4\,517\,117\,896\,542\,451\,200\,y_{1}^{7}y_{2}^{7}y_{3}^{7}y_{4}^{3},\:524\,002\,848\,759\,990\,600\,y_{1}^{12}y_{2}^{4}y_{3}^{4}y_{4}^{4},\:2\,305\,612\,534\,543\,958\,640\,y_{1}^{10}y_{2}^{6}y_{3}^{4}y_{4}^{4},\hfill \\ & 
\:3\,705\,448\,716\,231\,362\,100\,y_{1}^{8}y_{2}^{8}y_{3}^{4}y_{4}^{4},\:6\,916\,837\,603\,631\,875\,920\,y_{1}^{8}y_{2}^{6}y_{3}^{6}y_{4}^{4},\:-5\,533\,469\,423\,264\,502\,720\,y_{1}^{9}y_{2}^{5}y_{3}^{5}y_{4}^{5},\hfill \\ & 
\:-9\,485\,947\,582\,739\,147\,520\,y_{1}^{7}y_{2}^{7}y_{3}^{5}y_{4}^{5},\:12\,911\,430\,193\,446\,168\,384\,y_{1}^{6}y_{2}^{6}y_{3}^{6}y_{4}^{6}\hfill \end{smallmatrix}$

\vspace*{\fill}
\mbox{}

\noindent $\begin{smallmatrix}
f_{30}: &
\:96\,076\,794\,555\,968\,173\,y_{1}^{30},
\:467\,077\,693\,875\,y_{1}^{28}y_{2}^{2},
\:29\,425\,894\,714\,125\,y_{1}^{26}y_{2}^{4},
\:637\,561\,052\,139\,375\,y_{1}^{24}y_{2}^{6},
\hfill \\ & \:6\,284\,530\,371\,088\,125\,y_{1}^{22}y_{2}^{8},
\:32\,260\,589\,238\,252\,375\,y_{1}^{20}y_{2}^{10},
\:92\,871\,393\,261\,635\,625\,y_{1}^{18}y_{2}^{12},
\:156\,146\,408\,450\,881\,875\,y_{1}^{16}y_{2}^{14},
\hfill \\ & \:58\,851\,789\,428\,250\,y_{1}^{26}y_{2}^{2}y_{3}^{2},
\:3\,187\,805\,260\,696\,875\,y_{1}^{24}y_{2}^{4}y_{3}^{2},
\:58\,655\,616\,796\,822\,500\,y_{1}^{22}y_{2}^{6}y_{3}^{2},
\hfill \\ & \:483\,908\,838\,573\,785\,625\,y_{1}^{20}y_{2}^{8}y_{3}^{2},
\:2\,043\,170\,651\,755\,983\,750\,y_{1}^{18}y_{2}^{10}y_{3}^{2},
\:4\,736\,441\,056\,343\,416\,875\,y_{1}^{16}y_{2}^{12}y_{3}^{2},
\hfill \\ & \:6\,245\,856\,338\,035\,275\,000\,y_{1}^{14}y_{2}^{14}y_{3}^{2},
\:146\,639\,041\,992\,056\,250\,y_{1}^{22}y_{2}^{4}y_{3}^{4},
\:2\,258\,241\,246\,677\,666\,250\,y_{1}^{20}y_{2}^{6}y_{3}^{4},
\hfill \\ & \:15\,323\,779\,888\,169\,878\,125\,y_{1}^{18}y_{2}^{8}y_{3}^{4},
\:52\,100\,851\,619\,777\,585\,625\,y_{1}^{16}y_{2}^{10}y_{3}^{4},
\:94\,728\,821\,126\,868\,337\,500\,y_{1}^{14}y_{2}^{12}y_{3}^{4},
\hfill \\ & \:28\,604\,389\,124\,583\,772\,500\,y_{1}^{18}y_{2}^{6}y_{3}^{6},
\:156\,302\,554\,859\,332\,756\,875\,y_{1}^{16}y_{2}^{8}y_{3}^{6},
\:416\,806\,812\,958\,220\,685\,000\,y_{1}^{14}y_{2}^{10}y_{3}^{6},
\hfill \\ & \:574\,688\,181\,503\,001\,247\,500\,y_{1}^{12}y_{2}^{12}y_{3}^{6},
\:669\,868\,092\,254\,283\,243\,750\,y_{1}^{14}y_{2}^{8}y_{3}^{8},
\:1\,354\,622\,142\,114\,217\,226\,250\,y_{1}^{12}y_{2}^{10}y_{3}^{8},
\hfill \\ & \:1\,986\,779\,141\,767\,518\,598\,500\,y_{1}^{10}y_{2}^{10}y_{3}^{10},
\:-8\,718\,783\,602\,760\,y_{1}^{27}y_{2}y_{3}y_{4},
\:-1\,020\,097\,681\,522\,920\,y_{1}^{25}y_{2}^{3}y_{3}y_{4},
\hfill \\ & \:-30\,602\,930\,445\,687\,600\,y_{1}^{23}y_{2}^{5}y_{3}y_{4},
\:-368\,692\,447\,750\,426\,800\,y_{1}^{21}y_{2}^{7}y_{3}y_{4},
\:-2\,150\,705\,945\,210\,823\,000\,y_{1}^{19}y_{2}^{9}y_{3}y_{4},
\hfill \\ & \:-6\,686\,740\,302\,382\,740\,600\,y_{1}^{17}y_{2}^{11}y_{3}y_{4},
\:-11\,658\,931\,809\,282\,727\,200\,y_{1}^{15}y_{2}^{13}y_{3}y_{4},
\:-102\,009\,768\,152\,292\,000\,y_{1}^{23}y_{2}^{3}y_{3}^{3}y_{4},
\hfill \\ & \:-2\,580\,847\,134\,252\,987\,600\,y_{1}^{21}y_{2}^{5}y_{3}^{3}y_{4},
\:-25\,808\,471\,342\,529\,876\,000\,y_{1}^{19}y_{2}^{7}y_{3}^{3}y_{4},
\:-122\,590\,238\,877\,016\,911\,000\,y_{1}^{17}y_{2}^{9}y_{3}^{3}y_{4},
\hfill \\ & \:-303\,132\,227\,041\,350\,907\,200\,y_{1}^{15}y_{2}^{11}y_{3}^{3}y_{4},
\:-408\,062\,613\,324\,895\,452\,000\,y_{1}^{13}y_{2}^{13}y_{3}^{3}y_{4},
\hfill \\ &\:-54\,197\,789\,819\,312\,739\,600\,y_{1}^{19}y_{2}^{5}y_{3}^{5}y_{4},
\:-441\,324\,859\,957\,260\,879\,600\,y_{1}^{17}y_{2}^{7}y_{3}^{5}y_{4},
\hfill \\ &\:-1\,667\,227\,248\,727\,429\,989\,600\,y_{1}^{15}y_{2}^{9}y_{3}^{5}y_{4},
\:-3\,182\,888\,383\,934\,184\,525\,600\,y_{1}^{13}y_{2}^{11}y_{3}^{5}y_{4},
\hfill \\ &\:-2\,858\,103\,854\,961\,308\,553\,600\,y_{1}^{15}y_{2}^{7}y_{3}^{7}y_{4},
\:-8\,336\,136\,243\,637\,149\,948\,000\,y_{1}^{13}y_{2}^{9}y_{3}^{7}y_{4},
\hfill \\ &\:-11\,822\,156\,854\,612\,685\,380\,800\,y_{1}^{11}y_{2}^{11}y_{3}^{7}y_{4},
\:-18\,061\,628\,527\,880\,491\,554\,000\,y_{1}^{11}y_{2}^{9}y_{3}^{9}y_{4},
\hfill \\ &\:19\,126\,831\,564\,181\,250\,y_{1}^{24}y_{2}^{2}y_{3}^{2}y_{4}^{2},
\:879\,834\,251\,952\,337\,500\,y_{1}^{22}y_{2}^{4}y_{3}^{2}y_{4}^{2},
\hfill \\ &\:13\,549\,447\,480\,065\,997\,500\,y_{1}^{20}y_{2}^{6}y_{3}^{2}y_{4}^{2},
\:91\,942\,679\,329\,019\,268\,750\,y_{1}^{18}y_{2}^{8}y_{3}^{2}y_{4}^{2},
\hfill \\ &\:312\,605\,109\,718\,665\,513\,750\,y_{1}^{16}y_{2}^{10}y_{3}^{2}y_{4}^{2},
\:568\,372\,926\,761\,210\,025\,000\,y_{1}^{14}y_{2}^{12}y_{3}^{2}y_{4}^{2},
\hfill \\ &\:33\,873\,618\,700\,164\,993\,750\,y_{1}^{20}y_{2}^{4}y_{3}^{4}y_{4}^{2},
\:429\,065\,836\,868\,756\,587\,500\,y_{1}^{18}y_{2}^{6}y_{3}^{4}y_{4}^{2},
\hfill \\ &\:2\,344\,538\,322\,889\,991\,353\,125\,y_{1}^{16}y_{2}^{8}y_{3}^{4}y_{4}^{2},
\:6\,252\,102\,194\,373\,310\,275\,000\,y_{1}^{14}y_{2}^{10}y_{3}^{4}y_{4}^{2},
\hfill \\ &\:8\,620\,322\,722\,545\,018\,712\,500\,y_{1}^{12}y_{2}^{12}y_{3}^{4}y_{4}^{2},
\:4\,376\,471\,536\,061\,317\,192\,500\,y_{1}^{16}y_{2}^{6}y_{3}^{6}y_{4}^{2},
\hfill \\ &\:18\,756\,306\,583\,119\,930\,825\,000\,y_{1}^{14}y_{2}^{8}y_{3}^{6}y_{4}^{2},
\:37\,929\,419\,979\,198\,082\,335\,000\,y_{1}^{12}y_{2}^{10}y_{3}^{6}y_{4}^{2},
\hfill \\ &\:60\,957\,996\,395\,139\,775\,181\,250\,y_{1}^{12}y_{2}^{8}y_{3}^{8}y_{4}^{2},
\:89\,405\,061\,379\,538\,336\,932\,500\,y_{1}^{10}y_{2}^{10}y_{3}^{8}y_{4}^{2},
\hfill \\ &\:-8\,602\,823\,780\,843\,292\,000\,y_{1}^{21}y_{2}^{3}y_{3}^{3}y_{4}^{3},
\:-180\,659\,299\,397\,709\,132\,000\,y_{1}^{19}y_{2}^{5}y_{3}^{3}y_{4}^{3},
\hfill \\ &\:-1\,471\,082\,866\,524\,202\,932\,000\,y_{1}^{17}y_{2}^{7}y_{3}^{3}y_{4}^{3},
\:-5\,557\,424\,162\,424\,766\,632\,000\,y_{1}^{15}y_{2}^{9}y_{3}^{3}y_{4}^{3},
\hfill \\ &\:-10\,609\,627\,946\,447\,281\,752\,000\,y_{1}^{13}y_{2}^{11}y_{3}^{3}y_{4}^{3},
\:-3\,089\,274\,019\,700\,826\,157\,200\,y_{1}^{17}y_{2}^{5}y_{3}^{5}y_{4}^{3},
\hfill \\ &\:-20\,006\,726\,984\,729\,159\,875\,200\,y_{1}^{15}y_{2}^{7}y_{3}^{5}y_{4}^{3},
\:-58\,352\,953\,705\,460\,049\,636\,000\,y_{1}^{13}y_{2}^{9}y_{3}^{5}y_{4}^{3},
\hfill \\ &\:-82\,755\,097\,982\,288\,797\,665\,600\,y_{1}^{11}y_{2}^{11}y_{3}^{5}y_{4}^{3},
\:-100\,033\,634\,923\,645\,799\,376\,000\,y_{1}^{13}y_{2}^{7}y_{3}^{7}y_{4}^{3},
\hfill \\ &\:-216\,739\,542\,334\,565\,898\,648\,000\,y_{1}^{11}y_{2}^{9}y_{3}^{7}y_{4}^{3},
\:-331\,129\,856\,344\,475\,678\,490\,000\,y_{1}^{9}y_{2}^{9}y_{3}^{9}y_{4}^{3},
\hfill \\ &\:1\,072\,664\,592\,171\,891\,468\,750\,y_{1}^{18}y_{2}^{4}y_{3}^{4}y_{4}^{4},
\:10\,941\,178\,840\,153\,292\,981\,250\,y_{1}^{16}y_{2}^{6}y_{3}^{4}y_{4}^{4},
\hfill \\ &\:46\,890\,766\,457\,799\,827\,062\,500\,y_{1}^{14}y_{2}^{8}y_{3}^{4}y_{4}^{4},
\:94\,823\,549\,947\,995\,205\,837\,500\,y_{1}^{12}y_{2}^{10}y_{3}^{4}y_{4}^{4},
\hfill \\ &\:87\,529\,430\,721\,226\,343\,850\,000\,y_{1}^{14}y_{2}^{6}y_{3}^{6}y_{4}^{4},
\:284\,470\,649\,843\,985\,617\,512\,500\,y_{1}^{12}y_{2}^{8}y_{3}^{6}y_{4}^{4},
\hfill \\ &\:417\,223\,619\,771\,178\,905\,685\,000\,y_{1}^{10}y_{2}^{10}y_{3}^{6}y_{4}^{4},
\:670\,537\,960\,346\,537\,526\,993\,750\,y_{1}^{10}y_{2}^{8}y_{3}^{8}y_{4}^{4},
\hfill \\ &\:-42\,014\,126\,667\,931\,235\,737\,920\,y_{1}^{15}y_{2}^{5}y_{3}^{5}y_{4}^{5},
\:-210\,070\,633\,339\,656\,178\,689\,600\,y_{1}^{13}y_{2}^{7}y_{3}^{5}y_{4}^{5},
\hfill \\ &\:-455\,153\,038\,902\,588\,387\,160\,800\,y_{1}^{11}y_{2}^{9}y_{3}^{5}y_{4}^{5},
\:-780\,262\,352\,404\,437\,235\,132\,800\,y_{1}^{11}y_{2}^{7}y_{3}^{7}y_{4}^{5},
\hfill \\ &\:-1\,192\,067\,482\,840\,112\,442\,564\,000\,y_{1}^{9}y_{2}^{9}y_{3}^{7}y_{4}^{5},
\:531\,011\,879\,708\,773\,152\,690\,000\,y_{1}^{12}y_{2}^{6}y_{3}^{6}y_{4}^{6},
\hfill \\ &\:1\,251\,670\,859\,313\,536\,717\,055\,000\,y_{1}^{10}y_{2}^{8}y_{3}^{6}y_{4}^{6},
\:2\,011\,613\,881\,039\,612\,580\,981\,250\,y_{1}^{8}y_{2}^{8}y_{3}^{8}y_{4}^{6},
\hfill \\ &\:-2\,043\,544\,256\,297\,335\,615\,824\,000\,y_{1}^{9}y_{2}^{7}y_{3}^{7}y_{4}^{7}
\hfill \end{smallmatrix} $
\end{appendix}
\newcommand{\arxiv}[1]{}
\bibliographystyle{amsalpha} 
\bibliography{biblio.bib}

\end{document}